\documentclass[runningheads]{llncs}
\usepackage{xspace}
\usepackage{graphicx}
\usepackage{multirow}
\usepackage{amsmath}
\usepackage{amssymb}
\usepackage{fancyhdr}

\usepackage{xcolor}
\usepackage{booktabs}
\usepackage{hyperref}
\usepackage{marvosym}
\hypersetup{
    colorlinks=true, 
    linkcolor=blue,  
    filecolor=blue,  
    urlcolor=blue,   
    citecolor=blue,  
}
\newcommand{\our}{{MoreStyle}\xspace}
\definecolor{red}{RGB}{255,0,0}
\renewcommand\arraystretch{1.1}

\begin{document}
\title{MoreStyle: Relax Low-frequency Constraint of Fourier-based Image Reconstruction in Generalizable Medical Image Segmentation}



\author{
Haoyu Zhao\inst{1} \and 
Wenhui Dong\inst{1} \and 
Rui Yu\inst{2} \and 
Zhou Zhao\inst{3} \and \\ 
Du Bo\inst{1} \and
Yongchao Xu\inst{1}\textsuperscript{(\Letter)}
}

\institute{
    School of Computer Science, Wuhan University, Hubei, China \\ 
    \email{yongchao.xu@whu.edu.cn} \and
    University of Louisville, Louisville, USA \and
    School of Computer Science, Central China Normal University, \\ Hubei, China 
}

%
%
%
%
\maketitle              
\begin{abstract}
The task of single-source domain generalization (SDG) in medical image segmentation is crucial due to frequent domain shifts in clinical image datasets. To address the challenge of poor generalization across different domains, we introduce a Plug-and-Play module for data augmentation called \our. \our diversifies image styles by relaxing low-frequency constraints in Fourier space, guiding the image reconstruction network. With the help of adversarial learning, \our further expands the style range and pinpoints the most intricate style combinations within latent features. To handle significant style variations, we introduce an uncertainty-weighted loss. This loss emphasizes hard-to-classify pixels resulting only from style shifts while mitigating true hard-to-classify pixels in both \our-generated and original images. Extensive experiments on two widely used benchmarks demonstrate that the proposed \our effectively helps to achieve good domain generalization ability, and has the potential to further boost the performance of some state-of-the-art SDG methods. Source code is available at \url{https://github.com/zhaohaoyu376/morestyle}.

\end{abstract}

\section{Introduction}
Deep learning has significantly advanced medical image segmentation but faces challenges when applied across diverse domains. Varying imaging characteristics across healthcare centers, resulting from equipment discrepancies, operator skill levels, and considerations like radiation exposure and imaging time~\cite{guan2021domain,xie2021survey} leads to images of varied styles, which hinder deploying models in clinical settings.

To address these aforementioned challenges, many works explore unsupervised domain adaptation (UDA)~\cite{hu2022domain,wilson2020survey,wang2024advancing,wang2024dual} and multi-source domain generalization (MDG)~\cite{liu2020ms,wang2022generalizing,hu2022tmi,chen2023treasure}. UDA typically requires access to data from the source domain and the unlabeled target domain. Whereas, MDG necessitates access to data from multiple source domains. In clinical practice, acquiring target-domain data and redistributing multi-source domain data pose significant challenges due to high costs, privacy, and ethical concerns. Additionally, collecting and annotating extensive datasets is time-consuming, expensive, and sometimes impractical.

\begin{figure}[t]
  \centering
  \begin{minipage}{0.18\linewidth}
    \centering
    \includegraphics[width=\linewidth]{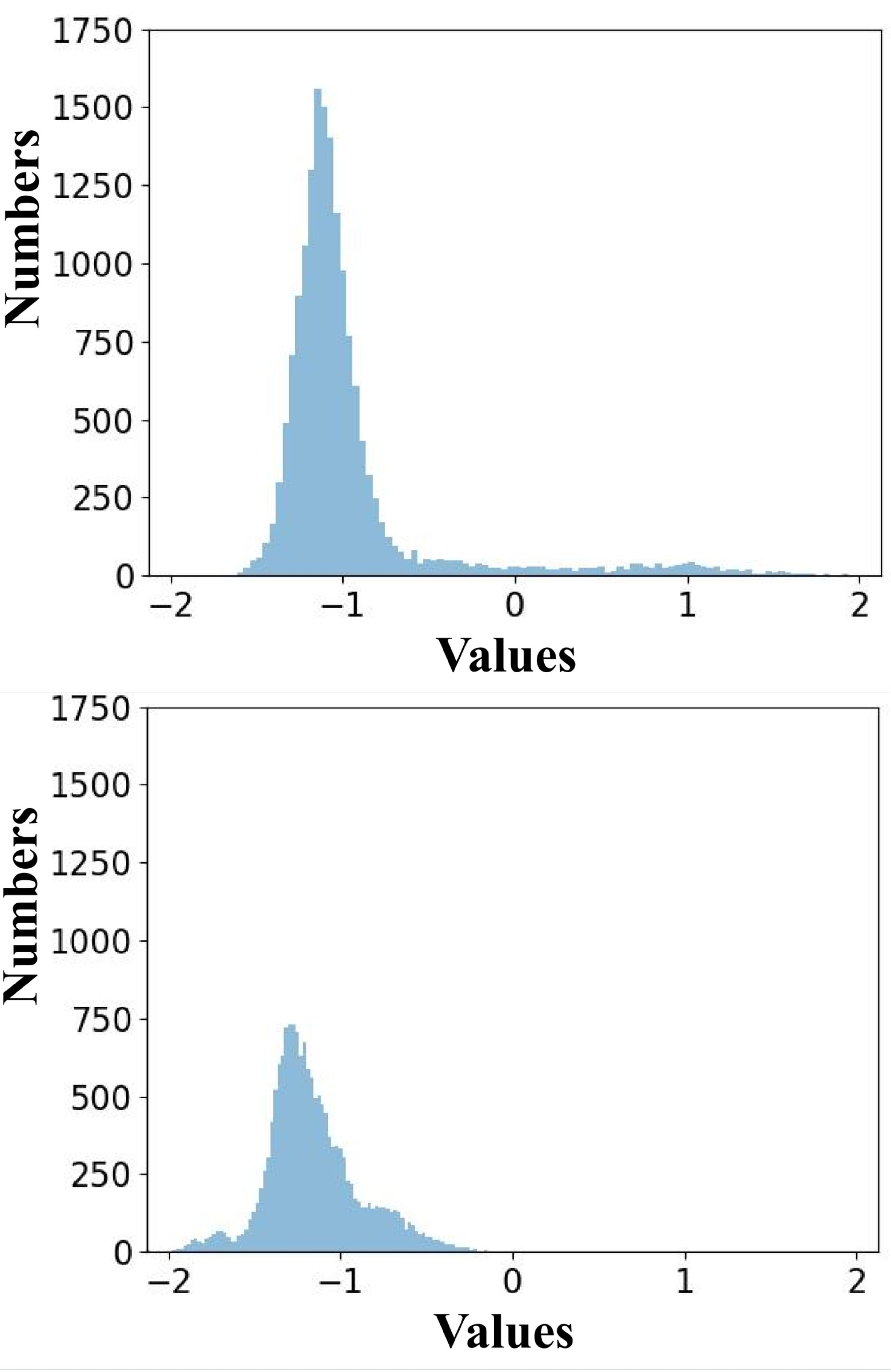}\\
    (a)
  \end{minipage}
  \hfill 
  \begin{minipage}{0.377\linewidth}
    \centering
    \includegraphics[width=\linewidth]{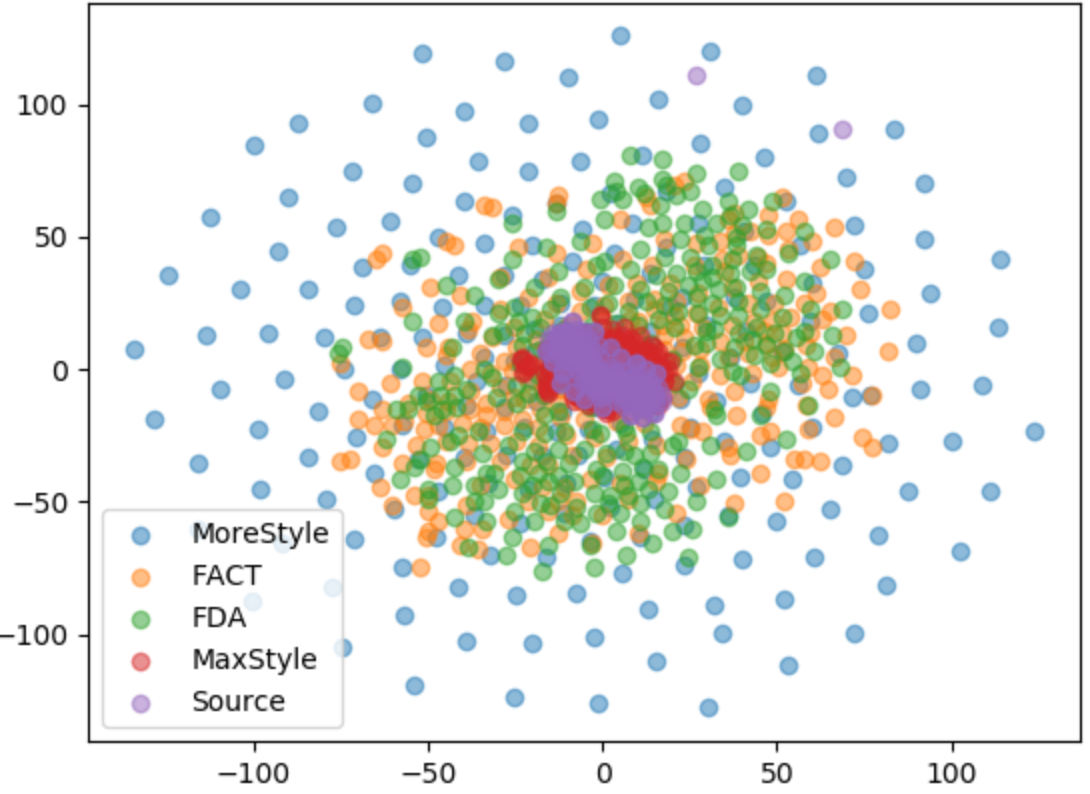}\\
    (b)
  \end{minipage}
  \hfill 
  \begin{minipage}{0.287\linewidth}
    \centering
    \includegraphics[width=\linewidth]{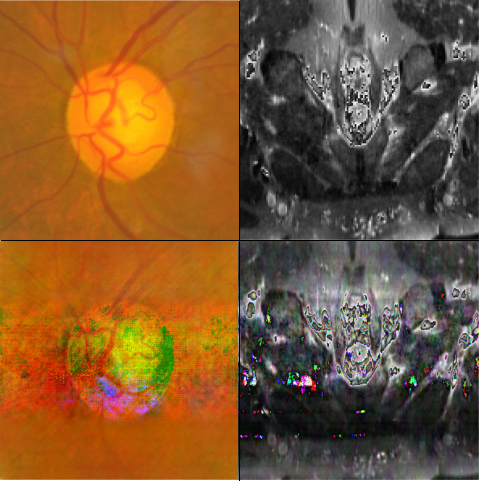}\\
    (c)
  \end{minipage}
  \caption{We adopt the first residual block of pre-trained ResNet-18~\cite{he2016deep} to extract image features. (a) Normalized feature distribution on the source images (the upper) and images generated by proposed \our (the lower). (b) Visualization of feature distribution on the source images and generated images given by methods including MaxStyle~\cite{chen2022maxstyle}, FDA~\cite{yang2020fda}, FACT~\cite{xu2021fourier}, and \our on RIGA+ dataset~\cite{almazroa2018retinal,hu2022domain,decenciere2014feedback} with tSNE. (c) Visualization of source image (the upper) and style-augmented images generated by proposed \our (the lower).}
  \label{fig:wia}
\end{figure}

Single-source domain generalization (SDG) is a more realistic yet challenging setting where a model is trained exclusively on data from one source domain and then deployed to segment data in unseen target domains. Various methods~\cite{cugu2022attention,hu2023devil,li2023frequency} have been introduced. A straightforward approach to enhance domain robustness is performing data augmentation at the image level~\cite{chen2022maxstyle,su2023rethinking,ouyang2022causality,xu2022adversarial}, which expands the range of the data and constrains the decision boundaries. Many Fourier-based data augmentation methods, such as FDA~\cite{yang2020fda}, FACT~\cite{xu2021fourier}, and others~\cite{wang2022domain,guo2023aloft,yang2020phase} emerge. They mainly perform Fourier transformations on images and exchange/mixup~\cite{zhang2017mixup} the low-frequency components of the amplitude to generate style-augmented images. Though randomly exchanging the low-frequency components of the amplitude effectively generates some images, the augmented styles are still very close to that of the source domain. The images generated by our \our is widely spread across the tSNE space, as shown in Fig.~\ref{fig:wia} (b).



In this work, we introduce a novel Plug-and-Play module called \our to address the style shifts across domains. \our combines an auxiliary reconstruction decoder with an adversarial noise encoder which is to generate perturbations for the reconstruction decoder. This process is supervised by a novel loss in Fourier space to generate images with same structure and of varied styles, as shown in Fig.~\ref{fig:wia}. Furthermore, we propose an uncertainty-weighted loss based on intersection-union between segmentation results to better handle the segmentation of these varied styles, focusing more on \our-induced style shifts and less on inherently difficult pixels.

Our method offers three key contributions: 1) We novelly propose to relax low-frequency constraint in Fourier space between the output of an image reconstruction network and the original image.
2) We propose a customized uncertainty-weighted loss to better deal with hard-to-classify pixels arising from style shifts and pay less attention to the true hard-to-classify pixels for the segmentation network.
3) \our significantly outperforms some state-of-the-art methods on generalizable OC/OD and prostate segmentation.


\begin{figure}[t]
  \centering
  \includegraphics[width=0.95\linewidth]{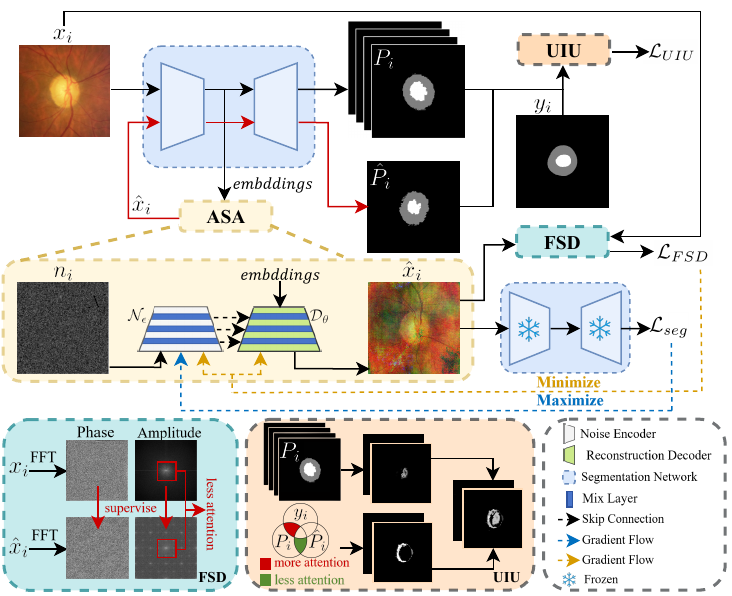}
   \caption{Pipeline of the proposed \our. In our proposed \our framework, we first generate style-augmented images $\hat{x}_i$ using Adversarial Style Augmentation (ASA) and a reconstruction decoder $\mathcal{D_\theta}$, with a noise encoder $\mathcal{N_\epsilon}$ generating style mixing $\gamma_i$ and noise perturbation $\beta_i$ through adversarial training to affect $\mathcal{D_\theta}$. This image reconstruction is guided by the Fourier Spectrum Diversity (FSD) loss $\mathcal{L}_{FSD}$. Lastly, both style-augmented $\hat{x}_i$ and source images $x_i$ are input into the segmentation network, supervised by a customized Uncertainty-weighted Intersection-Union (UIU) loss  $\mathcal{L}_{UIU}$.}
   \label{fig:pipline}
\end{figure}


\section{Method}
In this paper, to handle clinical data from unknown domains, we propose a novel Plug-and-Play module called \our as shown in Fig.~\ref{fig:pipline}. We employ an auxiliary reconstruction decoder to generate images of diverse styles, in parallel with an adversarial noise encoder which generates perturbations for the reconstruction decoder to fail the segmentation network. The source image $x_i$ and generated image $\hat{x}_i$ are utilized to compute Fourier Spectrum Diversity (FSD) loss $\mathcal{L}_{FSD}$ which is to ensure the diversity of the styles of generated images. Furthermore, we employ Uncertainty-weighted Intersection-Union (UIU) loss $\mathcal{L}_{UIU}$ to cope with the diverse styles of generated images $\{\hat{x}_i\}$.


\subsection{Adversarial Style Augmentation}
Adversarial Style Augmentation (ASA) focuses on generating data samples that fool the segmentation network to improve segmentation robustness. As shown in Fig.~\ref{fig:pipline}, ASA involves an auxiliary reconstruction decoder $\mathcal{D_\theta}$ and noise encoder $\mathcal{N_\epsilon}$. The noise encoder $\mathcal{N_\epsilon}$ generates high-dimensional style mixing $\gamma_i$ and noise perturbation $\beta_i$ from noise $n_i$ which is sampled from the Gaussian distribution. High-dimensional noise perturbation and style mixing disturbs the reconstruction decoder $\mathcal{D_\theta}$ to generate style-augmented images.

\begin{equation}
\begin{gathered}
 \hat{x}_i = \mathcal{D_\theta}{(x_i, \beta_i, \gamma_i)}, 
\label{eq:noise}
\end{gathered}
\end{equation}


where the image $x_i$ is transformed with style mixing $\gamma_i$ and noise perturbation $\beta_i$ into style augmented image $\hat{x}_i$.

\medskip
\noindent\textbf{Adversarial Training}.
ASA generates images of varied styles $\{\hat{x}_i\}$ by minimizing $\mathcal{L}_{FSD}$ and maximizing the segmentation loss $\mathcal{L}_{seg}$. Minimizing $\mathcal{L}_{FSD}$ updates both the noise encoder $\mathcal{N_\epsilon}$ and reconstruction decoder $\mathcal{D_\theta}$, encouraging to preserve the structure of the original image $x_i$ for the generated one $\hat{x}_i$. Maximizing $\mathcal{L}_{seg}$ updates only the noise encoder $\mathcal{N_\epsilon}$, forcing to generate more challenging style mixing $\gamma_i$ and noise perturbation $\beta_i$ to disturb the reconstruction decoder, which generates images that fool the segmentation network and thus improves the generalization ability of the segmentation network.


\subsection{Fourier Spectrum Diversity Loss}\label{sec:FSD}
Low-frequency components, containing most of the energy distributions, efficiently capture style variations across diverse domains. In contrast, high-frequency components focus on object structures resembling identity~\cite{xu2021fourier,wang2022domain,1456290}. Therefore, it is reasonable for the reconstruction decoder $\mathcal{D_\theta}$ to pay less attention to the low-frequency components to generate images of diverse styles. To this end, we propose a novel reconstruction loss called Fourier Spectrum Diversity (FSD) loss $\mathcal{L}_{FSD}$ which relaxes the low-frequency constraint in Fourier space to encourage the reconstruction decoder $\mathcal{D_\theta}$ to generate images with more diverse low-frequency components, thereby also enhancing the stylistic diversity of the images.

For an image $x_{i}\in\mathbb{R}^{H\times W\times C}$, where $H$, $W$, $C$ denote the height, width and number of channels, respectively, its frequency space signal $\mathcal{F}(x_i)$ can be obtained with Fast Fourier Transform (FFT), which is defined as follows:

\begin{equation}
\begin{aligned}
\mathcal{F}(x_i)(u,v,c) & =\sum_{h=1}^H\sum_{w=1}^Wx_i(h,w,c)e^{-j2\pi(\frac hHu+\frac wWv)} =\mathcal{A}(x_i)e^{j\mathcal{P}(x_i)},
\end{aligned}
\label{eq:fft}
\end{equation}

where $j^2 = -1$, and $\mathcal{A}(x_i)$ and $\mathcal{P}(x_i)$ refer to the amplitude and phase spectra of $x_i$, respectively.
We center the low-frequency components within the frequency spectrum, and then introduce a binary mask $\mathcal{M}\in\mathbb{R}^{H\times W}$, where all values are zero except in the central region.
Following~\cite{hu2023devil}, the low-frequency components $\mathcal{A}_l(x_i)$ and high-frequency components $\mathcal{A}_h(x_i)$ are given by:

\begin{equation}
\begin{gathered}
\mathcal{A}_l(x_i)=\mathcal{M}\odot\mathcal{A}(x_i), \\
\mathcal{A}_h(x_i)=(I-\mathcal{M})\odot\mathcal{A}(x_i),\end{gathered}
\label{eq:high_low}
\end{equation}

where $\odot$ denotes element-wise multiplication. The Fourier Spectrum Diversity Loss $\mathcal{L}_{FSD}$ is then defined as follows:

\begin{equation}
\begin{aligned}
&\mathcal{L}_{FSD}(x_i,\hat{x}_i) = \mathcal{L}_{mse}(\mathcal{P}(x_i),\mathcal{P}(\hat{x}_i))+\lambda_1\mathcal{L}_{mse}(x_i,\hat{x}_i)
\\
&+\lambda_2\mathcal{L}_{mse}(\mathcal{A}_h(x_i),\mathcal{A}_h(\hat{x}_i))
+\lambda_3\mathcal{L}_{mse}(\mathcal{A}_l(x_i),\mathcal{A}_l(\hat{x}_i)),
\end{aligned}
\label{eq:fsd}
\end{equation}

where $\lambda_1$, $\lambda_2$ and $\lambda_3$ decay exponentially at a rate of 0.99 raised to the power of the epoch number, thereby encouraging $\mathcal{D_\theta}$ to generate images of more styles.



\subsection{Uncertainty-weighted Intersection-Union Loss}
It is crucial to effectively leverage these style-augmented images, therefore, we propose an Uncertainty-weighted Intersection-Union (UIU) loss $\mathcal{L}_{UIU}$ between the segmentation results of original image $x_i$ and generated style-augmented image $\hat{x}_i$. We pay more attention to the hard-to-classify pixels arising only from style shifts, while giving less emphasis to the common hard-to-classify pixels between the images generated above and the original one.

We identify challenging pixels due to style shifts which are those easily segmented in $x_i$ but problematic in $\hat{x}_i$. Our aim is to make the network focus more on the hard-to-classify pixels arising only from style shifts $P_i^k \cap y_i^k - P_i^k \cap y_i^k \cap \hat{P}_i^k$ (more attention region in Fig.~\ref{fig:pipline}), where k means the k-th class. The collection of these pixels is denoted as $m_i$.
The common hard-to-classify pixels between the images generated above and the original one are $P_i^k \cap \hat{P}_i^k - P_i^k \cap y_i^k \cap \hat{P}_i^k$ (less attention regions in Fig.~\ref{fig:pipline}). These pixels are grouped into a set called $l_i$.

We then estimate pixel-wise uncertainty using the Monte Carlo Dropout method~\cite{kendall2017uncertainties,gal2016dropout} to find pixels that pose challenge to network segmentation. We perform T stochastic forward passes through segmentation network using $x_i^T$ which is generated by adding random Gaussian noise on $x_i$. The uncertainty map $u_i$ is created by calculating the entropy of the predictions, highlighting pixels with high entropy as the most challenging ones. We do not consider $\hat{x}_i$ in this step because it will lead the uncertainty map to emphasize challenging pixels due to immense style variations rather than inherently challenging pixels.







Given uncertainty map $u_i$ and two sets of pixels $m_i$ and $l_i$, for every pixel in $m_i$, we enhance its value in $u_i$ by applying a multiplication factor of $\rho$. Conversely, for pixels within $l_i$, their values in $u_i$ are adjusted by multiplying with a factor of $\sigma$. We set $\rho$ to 1.2 and $\sigma$ to 0.8 in this paper. Finally, the Uncertainty-weighted Intersection-Union loss $\mathcal{L}_{UIU}$ for $i$-th image is given by:


\begin{equation}
\begin{aligned}
\mathcal{L}_{UIU}(P_i,y_i) = & -\frac{1}{HW}\sum_{j=1}^{HW}u_i^j(y_i^j\log P_i^j + (1-y_i^j)\log(1-P_i^j)).
\end{aligned}
\label{eq:uce}
\end{equation}

The final segmentation loss is defined as $\mathcal{L}_{seg} = \mathcal{L}_{UIU}(P_i,y_i) + \mathcal{L}_{dice}(P_i,y_i)$.

\textbf{\section{Experiments}}
\subsection{Dataset and Training Details}
We conduct experiments on two public datasets the RIGA+ dataset\footnote{\url{https://zenodo.org/records/6325549/}}~\cite{almazroa2018retinal,hu2022domain,decenciere2014feedback} and the prostate dataset\footnote{\url{http://medicaldecathlon.com/}}~\cite{liu2020ms}. In RIGA+ dataset, each image is resized to 512$\times$512 pixels, and we select BinRushed and Magrabia as the source domains to train our segmentation network. In prostate dataset, we preprocess these MRI cases following the methods of a prior study~\cite{hu2022domain}, and then resize slices to 384$\times$384 pixels with the same voxel spacing. To evaluate the segmentation performance, we employ the Dice Similarity Coefficient.

In all our experiments, we use the same backbone data augmentation strategy as in~\cite{hu2023devil}. We set $\lambda_2$ to $5\times10^{-5}$, $\lambda_3$ to $5\times10^{-6}$ and $\lambda_1$ to $5\times10^{-3}$ in Eq.~\ref{eq:fsd}. In the reconstruction decoder $\mathcal{D_\theta}$, mix layers are integrated within the first three convolutional blocks and are activated at a probability of 0.5 to introduce feature perturbation. For the noise encoder $\mathcal{N_\epsilon}$, we employ a specific noise optimizer, whereas another optimizer is utilized for both $\mathcal{D_\theta}$ and $\mathcal{N_\epsilon}$. We set learning rate to $1\times10^{-2}$ for segmentation network. For MedSAM~\cite{ma2024segment}, the learning rate is set to $1\times10^{-5}$ on RIGA+ and $1\times10^{-7}$ on prostate. All experiments are conducted using the PyTorch framework on an NVIDIA 3090 GPU.

\begin{table*}[t]
\centering
\caption{The experimental results on OC/OD segmentation. The source domains are BinRushed (top 2 groups) and Magrabia (bottom 2 groups), respectively. The target domains are BASE1, BASE2, and BASE3. The best result is in \textcolor{red}{red}.}
\scriptsize
\setlength{\tabcolsep}{5pt}
\renewcommand{\arraystretch}{1} 
\begin{tabular}{l|c|c|c|c|c|c|c|c}
\hline
\multirow{2}{*}{Method} & \multicolumn{2}{c|}{Base1} & \multicolumn{2}{c|}{Base2} & \multicolumn{2}{c|}{Base3} & \multicolumn{2}{c}{Avg} \\
\cline{2-9}
 & OC & OD & OC & OD & OC & OD & OC & OD \\
 \hline
MixStyle \tiny{(ICLR2021)}~\cite{zhou2021domain}    & 92.59 & 81.12 & 89.32 & 72.90 & 93.67 & 78.62 & 91.84 & 77.70\\
DSU \tiny{(ICLR2022)}~\cite{li2022uncertainty}         & 91.70 & 82.63 & 82.54 & 70.26 & 90.15 & 77.97 & 88.26 & 77.23\\
EFDM \tiny{(CVPR2022)}~\cite{zhang2022exact}         & 93.32 & 78.20 & 93.20 & 79.78 & 91.34 & 80.63 & 92.70 & 80.63\\
MaxStyle \tiny{(MICCAI2022)}~\cite{chen2022maxstyle} & 95.22 & 82.49 & 94.46 & 79.19 & 93.96 & 79.27 & 94.60 & 80.47 \\
SLAug \tiny{(AAAI2023)}~\cite{su2023rethinking}          & 93.17 & 81.18 & 91.60 & 77.32 & 94.77 & 81.96 & 93.13 & 80.15\\
MoreStyle \tiny{(Ours)} & \textcolor{red}{96.08} & 85.58 & \textcolor{red}{96.20} & \textcolor{red}{88.29} & \textcolor{red}{96.03} & \textcolor{red}{87.11} & \textcolor{red}{96.10} & \textcolor{red}{86.91}\\
\hline
CCSDG  \tiny{(MICCAI2023)}~\cite{hu2023devil}         & 95.45 & 86.14 & 95.68 & 86.42 & 95.62 & 85.60 & 95.57 & 86.07\\
CCSDG+MoreStyle  & 95.70 & \textcolor{red}{86.30} & 96.08 & 86.91 & 95.55 & 86.06 & 95.78 & 86.42 \\
MedSAM \tiny{(Nat. Commun2024)}~\cite{ma2024segment} & 94.15 & 82.19 & 93.42 & 86.55 & 94.00 & 83.39 & 93.87 & 83.96\\
MedSAM+MoreStyle & 94.62 & 84.62 & 94.77 & 86.14 & 94.48 & 84.64 & 94.63 & 85.12\\
\hline
\hline
MixStyle \tiny{(ICLR2021)}~\cite{zhou2021domain}    & 93.76 & 80.26 & 92.93 & 78.65 & 90.94 & 82.63 & 92.66 & 80.43 \\
DSU  \tiny{(ICLR2022)}~\cite{li2022uncertainty}         & 93.65 & 81.62 & 92.09 & 77.63 & 91.40 & 82.57 & 92.48 & 80.60 \\
EFDM  \tiny{(CVPR2022)}~\cite{zhang2022exact}         & 93.10 & 80.84 & 91.11 & 78.54 & 91.70 & 82.23 & 92.04 & 80.50 \\
MaxStyle \tiny{(MICCAI2022)}~\cite{chen2022maxstyle}  & 94.54 & 83.24 & 94.64 & 85.27 & 93.99 & 84.67 & 94.41 & 84.32 \\
SLAug  \tiny{(AAAI2023)}~\cite{su2023rethinking}          & 92.76 & 82.07 & 92.83 & 81.66 & 90.94 & 81.71 & 92.25 & 81.83\\
MoreStyle \tiny{(Ours)} & 95.04 & \textcolor{red}{87.38} & \textcolor{red}{95.63} & 87.71 & 95.25 & 86.16 & \textcolor{red}{95.29} & 87.13 \\
\hline
CCSDG  \tiny{(MICCAI2023)}~\cite{hu2023devil}         & \textcolor{red}{95.15} & 84.57 & 95.25 & 82.17 & 95.09 & 83.84 & 95.16 & 83.57 \\
CCSDG+MoreStyle  & 95.11 & 86.47 & 95.13 & \textcolor{red}{90.10} & \textcolor{red}{95.60} & \textcolor{red}{87.83} & 95.26 & \textcolor{red}{88.05} \\
MedSAM \tiny{(Nat. Commun2024)}~\cite{ma2024segment} & 93.32 & 69.56 & 93.96 & 77.49 & 93.84 & 77.36 & 93.68 & 74.44 \\
MedSAM+MoreStyle& 94.30 & 73.08 & 94.38 & 80.59 & 93.99 & 77.53 & 94.24 & 76.79 \\
\hline
\end{tabular}
\label{tab:ocod}
\end{table*}

\begin{table*}[t]
\centering
\caption{The experimental results on prostate segmentation. Each column is the Dice score averaged on the rest domains except the source domain for training.}
\scriptsize
\renewcommand{\arraystretch}{1} 
\setlength{\tabcolsep}{1pt}
\begin{tabular}{l|c|c|c|c|c|c|c} 
\hline
Method & A to Rest & B to Rest & C to Rest & D to Rest & E to Rest & F to Rest & Avg\\
\hline
MixStyle \tiny{(ICLR2021)}~\cite{zhou2021domain}       & 72.32 & 64.19 & 40.27 & 60.23 & 40.33 & 48.59 & 54.32 \\
DSU \tiny{(ICLR2022)}~\cite{li2022uncertainty}      & 73.51 & 64.85 & 45.08 & 64.05 & 42.27 & 43.38 & 55.52 \\
EFDM \tiny{(CVPR2022)}~\cite{zhang2022exact}         & 73.53 & 64.06 & 45.49 & 61.15 & 41.07 & 47.45 & 55.46 \\
MaxStyle \tiny{(MICCAI2022)}~\cite{chen2022maxstyle}       & 73.40 & 63.63 & 51.45 & 62.15 & 38.14 & 56.31 & 57.51 \\
SLAug  \tiny{(AAAI2023)}~\cite{su2023rethinking}         & 79.31 & 68.47 & 58.47 & 68.25 & 48.88 & 57.61 & 63.50\\
Morestyle \tiny{(Ours)}      & 80.06 & 74.42 & 54.16 & 65.96 & 51.12 & 54.19 & 63.32 \\
\hline
CCSDG (\tiny{MICCAI2023})~\cite{hu2023devil}        & 80.62 & 69.52 & 65.18 & 67.89 & 58.99 & 63.27 & 67.58 \\
CCSDG+MoreStyle    & 80.67 & 75.99 & 66.28 & 69.57 & 60.38 & 67.22 & 70.02\\
MedSAM\tiny{(Nat. Commun2024)}~\cite{ma2024segment}        & 88.02 & 87.92 & 87.21 & 88.11 & 87.42 & 88.43 & 87.85 \\
MedSAM+MoreStyle    & \textcolor{red}{88.43} & \textcolor{red}{88.69} & \textcolor{red}{87.47} & \textcolor{red}{88.50} & \textcolor{red}{88.32} & \textcolor{red}{88.59} & \textcolor{red}{88.33}\\
\hline
\end{tabular}
\label{tab:prostate} 
\end{table*}

\subsection{Comparison with State-of-the-art Methods}
We conduct comparative experiments against various state-of-the-art (SOTA) methods, including (1) MaxStyle~\cite{chen2022maxstyle} using adversarial noise; (2-4) MixStyle~\cite{zhou2021domain}, DSU~\cite{li2022uncertainty} and EFDM~\cite{zhang2022exact} with feature-space domain randomization; (5) SLAug~\cite{su2023rethinking} applying Bézier transformation to both global and local regions; (6) CCSDG~\cite{hu2023devil} finding invariant between domains; (7) Medical foundation model, MedSAM~\cite{ma2024segment}. 

\begin{figure*}[t]
  \centering
  \includegraphics[width=1\linewidth]{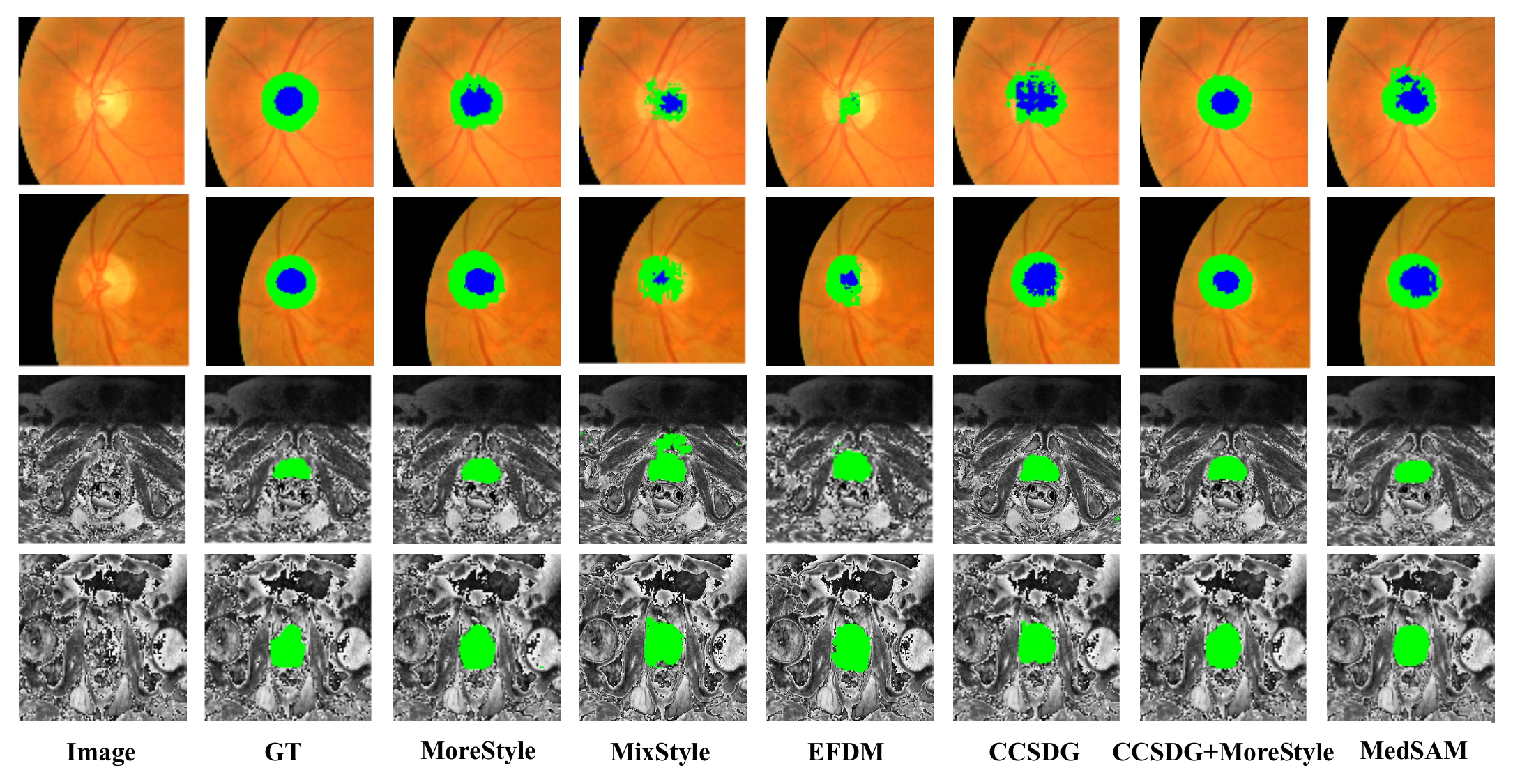} 
   \caption{Visualization of results by \our and several SOTA methods.}
   \label{fig:onecol}
\end{figure*}

As shown in Table~\ref{tab:ocod} and Table~\ref{tab:prostate}, \our yields good results on both datasets. Notably, in comparison with recent methods such as MedSAM~\cite{ma2024segment} and CCSDG~\cite{hu2023devil}, the application of \our yields notable improvement. Some qualitative results are shown in Fig.~\ref{fig:onecol}, compared with some SOTA methods, \our achieves more accurate segmentation result.

Unlike other adversarial data augmentation methods that may need long training times, \our achieves good results in only 100 epochs, significantly less than the 1500 epochs required by MaxStyle~\cite{chen2022maxstyle} for satisfactory outcomes.

\medskip
\noindent
\textbf{Ablation Studies.}
To evaluate the effectiveness of our proposed modules, including ASA, FSD and UIU, we conduct ablation experiments using RIGA+ dataset~\cite{almazroa2018retinal,hu2022domain,decenciere2014feedback}. 
The corresponding results are shown in Table~\ref{tab:ablation}. It reveals that ASA, FSD and UIU all contribute to performance gains. Additional hyper-parameter ablation studies are in supplementary materials.
\our increases the number of parameters by 8.6\% compared to the baseline. Considering the significant performance improvement over the baseline model \textbf{without any extra inference time}, this small amount of extra cost is deserving. 
It is worth mentioning that different from other DG methods, \our does not necessarily sacrifice the in-dataset performance, as shown in Table.~\ref{tab:ablation}.

\medskip
\noindent
\textbf{Comparison with classic Fourier-based data augmentation methods.}
ASA+FSD has much better performance than traditional Fourier-based data augmentation such as FDA~\cite{yang2020fda} and FACT~\cite{xu2021fourier}, as shown in Table~\ref{tab:compare_fourier} and supplementary materials. This indicates that, unlike direct swapping in the Fourier domain, our \our can generate images of more diverse styles, improving the generalization ability of segmentation network.

\begin{table}[t]
\centering
\begin{minipage}[t]{0.6\linewidth} 
\centering
\caption{Ablation studies on the impact of each module in our method under both in-dataset and cross-dataset settings.}
\scriptsize
\setlength{\tabcolsep}{1pt}
\renewcommand{\arraystretch}{0.95} 
\begin{tabular}{l|c|c|c}
\hline
\multirow{2}{*}{Method} & \multirow{2}{*}{\#Params} & \multicolumn{2}{c}{Average} \\
\cline{3-4}
 & & OC & OD \\
 \hline
 Baseline (cross-dataset) & 43.80M & 89.12 & 77.90\\
 Baseline+ASA (cross-dataset) & 47.56M & 93.11 & 81.63\\
 Baseline+ASA+FSD (cross-dataset) & 47.56M & 95.42 & 86.27 \\
 \our (cross-dataset) & 47.56M & \textcolor{red}{95.70} & \textcolor{red}{87.02} \\
  \hline
 Baseline (in-dataset) & 43.80M & 95.64 & 84.25\\
 \our(in-dataset) & 47.56M & \textcolor{red}{96.10  } & \textcolor{red}{85.29} \\
\hline
\end{tabular}
\label{tab:ablation}
\end{minipage}
\hfill
\begin{minipage}[t]{0.38\linewidth} 
\centering
\caption{Comparison between data augmentation in \our with some classical Fourier-based data augmentations.}
\scriptsize
\setlength{\tabcolsep}{3pt}
\renewcommand{\arraystretch}{0.95}
\begin{tabular}{l|c|c}
\hline
\multirow{2}{*}{Method} & \multicolumn{2}{c}{Average} \\
\cline{2-3}
 & OC & OD \\
 \hline
 Baseline & 89.12 & 77.90 \\
 Baseline+FDA~\cite{yang2020fda} & 94.33 & 83.81 \\
 Baseline+FACT~\cite{xu2021fourier} & 93.94 & 82.25 \\
 Baseline+ASA+FSD & \textcolor{red}{95.42} & \textcolor{red}{86.27}\\
\hline
\end{tabular}
\label{tab:compare_fourier}
\end{minipage}
\end{table}


\section{Conclusion}
In this paper, we propose a novel Plug-and-Play module called \our to significantly enhance the diversity of the training data by relaxing low-frequency constraint in Fourier space to supervise the reconstruction decoder. Combined with adversarial style augmentation, we effectively generate images of diverse styles. We then propose an uncertainty-weighted loss to better make use of these diversely style-augmented images. In particular, on one hand, we propose to pay more attention to hard-to-classify pixels arising only from domain shifts to explore the potential of generated images for segmentation network. On the other hand, we pay less attention to the true hard-to-classify pixels for both the generated and original images to cope with the diversely augmented styles. Extensive experiments on two public datasets demonstrate that the proposed \our is effective, significantly/consistently boosting the performance of some state-of-the-art methods. The limitation is that while \our excels in generating images of various styles, it struggles to generalize effectively to images with geometric variations. Integrating \our into geometric data augmentation would be an interesting direction.

{
    \small
    \bibliographystyle{splncs04}
    \bibliography{Paper-0782}
}

\end{document}